\documentclass[12pt,preprint]{aastex}
\usepackage{emulateapj5}
\usepackage{apjfonts}
\usepackage{epsfig}
\usepackage{natbib}

\newcommand{\cat}{CaT}

\newcommand{\chisq}{\ensuremath{\chi^2}}
\newcommand{\chisqr}{\ensuremath{\chi^2_r}}

\newcommand{\etal}{et al.}

\newcommand{\epsi}{\ensuremath{\epsilon_0}}

\def\gtrsim{\mathrel{\hbox{\rlap{\hbox{\lower4pt\hbox{$\sim$}}}\hbox{\raise2pt\hbox{$>$}}}}}
\newcommand{\fwha}{\ensuremath{\mathrm{FWHM}_\mathrm{H{\alpha}}}}

\newcommand{\halpha}{H\ensuremath{\alpha}}

\newcommand{\hbeta}{H\ensuremath{\beta}}

\newcommand{\kms}{km~s\ensuremath{^{-1}}}

\newcommand{\lf}{\ensuremath{L_{5100}}}
\newcommand{\lum}{ergs s$^{-1}$}

\newcommand{\lledd}{\ensuremath{L_{\mathrm{bol}}/L{\mathrm{_{Edd}}}}}

\newcommand{\lha}{\ensuremath{L_{\mathrm{H{\alpha}}}}}

\newcommand{\mbh}{\ensuremath{M_\mathrm{BH}}}

\newcommand{\msigma}{\ensuremath{M_{\mathrm{BH}}-\sigmastar}}
\newcommand{\msun}{\ensuremath{M_{\odot}}}

\newcommand{\nii}{[\ion{N}{2}]}

\newcommand{\rblr}{\ensuremath{R_{\mathrm{BLR}}}}

\newcommand{\sii}{[\ion{S}{2}]}

\newcommand{\sigha}{\ensuremath{\sigma_{\mathrm{H\alpha}}}}
\newcommand{\sigmastar}{\ensuremath{\sigma_{\ast}}}

\def\lax{{$\mathrel{\hbox{\rlap{\hbox{\lower4pt\hbox{$\sim$}}}\hbox{$<$}}}$}}
\def\gax{{$\mathrel{\hbox{\rlap{\hbox{\lower4pt\hbox{$\sim$}}}\hbox{$>$}}}$}}

\slugcomment{To appear in {\it The Astrophysical Journal (Letters)}.}
\shorttitle{The \msigma\ Relation for AGNs}
\shortauthors{GREENE \& HO}

\begin{document}

\title{The \msigma\ Relation in Local Active Galaxies}

\author{Jenny E. Greene}
\affil{Harvard-Smithsonian Center for Astrophysics, 60 Garden St., 
Cambridge, MA 02138}

\and

\author{Luis C. Ho}
\affil{The Observatories of the Carnegie Institution of Washington,
813 Santa Barbara St., Pasadena, CA 91101}

\begin{abstract}

We examine whether active galaxies obey the same relation between
black hole mass and stellar velocity dispersion as inactive systems,
using the largest published sample of velocity dispersions for active
nuclei to date.  The combination of 56 original measurements with
objects from the literature not only increases the sample from the 15
considered previously to 88 objects, but allows us to cover an
unprecedented range in both stellar velocity dispersion (30--268 \kms)
and black hole mass ($10^5-10^{8.6}$~\msun).  In the \msigma\
relation of active galaxies we find a lower zeropoint than the best-fit
relation of Tremaine et al. (2002) for inactive galaxies, and an intrinsic 
scatter of 0.4 dex.  There is also evidence for a flatter slope at low black 
hole masses.  We discuss potential contributors to the
observed offsets, including variations in the geometry of the broad-line 
region, evolution in the \msigma\ relation, and differential
growth between black holes and galaxy bulges.


\end{abstract}

\section{Introduction}

Black holes (BHs) are a basic component of galaxies, and the existence of a 
tight correlation between the stellar velocity dispersion of bulges 
(\sigmastar) and the
BH mass (the \msigma\ relation; Gebhardt \etal\ 2000a; Ferrarese \&
Merritt 2000) suggests that the growth of the BH plays a fundamental
role in the growth of the bulge, although exactly how remains unclear 
(e.g.,~Silk \& Rees 1998; Kauffmann \& Haehnelt 2000; Di~Matteo \etal\ 2005).  
The \msigma\ relation for inactive galaxies, in so far as it
represents the final state of the BH-bulge system, represents a
boundary condition for various evolutionary scenarios, and some clues
are embedded in the scatter and possible skewness in the local
\msigma\ relation (Robertson \etal\ 2005).  Unfortunately, the number
of available points in the relation for inactive galaxies is limited, and
statistics are poor.  Currently, our only recourse is to rely on BH
masses from active galactic nuclei (AGNs).  To the extent that the BHs
in AGNs continue to gain mass, the relation of BH masses in AGNs to
their host bulges may carry additional information about the
establishment of the relation for inactive sources.  Since the majority of BH
mass was assembled at high redshift (e.g.,~Yu \& Tremaine 2002), we
might expect to find the strongest evidence for evolution in the
\msigma\ relation at large cosmological distances (Shields \etal\
2003; Treu \etal\ 2004; Walter \etal\ 2004; Peng \etal\ 2005).  
While undoubtedly this
is a vital direction to pursue, there are a number of compelling
reasons to study the local AGN \msigma\ relation as well.

For one thing, while various methods are available to characterize the
bulge potential, virial mass estimation (e.g.,~Ho 1999; Wandel \etal\ 1999;
Kaspi \etal\ 2000), where the dense broad-line region (BLR) gas
is assumed to be on Keplerian orbits around the central BH, is currently
the most widely utilized tracer of BH mass.
In the absence of a detailed
model of the BLR, the zeropoint for the virial BH mass scale is set
through a direct comparison with stellar velocity dispersions for a
small sample of local AGNs (Gebhardt \etal\ 2000b; Ferrarese \etal\
2001; Nelson \etal\ 2004; Onken \etal\ 2004).  The virial BH masses
considered in these studies are remarkably consistent with the
\msigma\ relation of inactive galaxies, suggesting virial masses are
reliable.  But we must be cautious.  For instance, there are compelling
reasons to believe the BLR is actually a disk-wind (e.g.,~Murray \&
Chiang 1997; Proga et al. 2000; Proga \& Kallman 2004)
whose kinematics depend on both the mass and the accretion rate onto
the BH.  In this scenario, the virial mass calibration would depend
systematically on BH mass and luminosity.  We require objects spanning
a wide range of AGN properties to properly calibrate the primary
rung in our AGN BH ``mass ladder.''  At the same time, we may hope to
learn about evolution of the \msigma\ relation by looking at the full
distribution of local BHs (Robertson \etal\ 2005), and AGNs in
particular (e.g.,~Yu \& Lu 2004), in the \msigma\ plane.  

\section{Sample Selection and Methodology}

Our goal is to directly compare \mbh\ with \sigmastar\ for a large
sample of AGNs.  As outlined in Greene \& Ho (2005c), we selected
spectroscopically identified AGNs from the Third Data Release of the
Sloan Digital Sky Survey (SDSS DR3; Abazajian \etal\ 2005) with $z
\leq 0.05$ and signal-to-noise (S/N) ratios per pixel $\geq 18$ in the
region surrounding the \ion{Ca}{2} $\lambda \lambda$8498, 8542, 8662
triplet (CaT).  Here we consider the 32 objects from Greene \& Ho
(2005c) with resolved \sigmastar\ measurements and robust \halpha\
line widths.  To increase the sample, we have relaxed the $z$
requirement and included an additional 24 galaxies to satisfy the
criteria established by Greene \& Ho for use of the ``Fe region''
(5250--5820 \AA): S/N $\geq 20$ ($\langle$ S/N $\rangle=40 \pm 8$),
estimated Eddington ratios [\lledd, where $L_{\mathrm{Edd}} \equiv
1.26 \times 10^{38}$~(\mbh/\msun) \lum] $\lesssim 0.5$, and AGN
contamination $< 80 \%$.  
In addition to the 56 objects from the SDSS,
we include 15 intermediate-mass BHs (IMBHs; \mbh\ $\leq$ $10^6$~\msun)
from Greene \& Ho (2004) with \sigmastar\ measurements from Barth
\etal\ (2005), resulting in a total of 71 objects
(see Table 1).

The stellar velocity dispersions are measured using a 
\psfig{file=table1_sh_v4.epsi,width=9cm,angle=0} 
\vskip +3mm 
\noindent
direct pixel-fitting algorithm described in detail in Greene \& Ho
(2005c).  The uncertainties are derived from simulations in which the
AGN contamination and S/N ratio are varied for a grid of model galaxy spectra.

Virial masses are based on reverberation mapping (Blandford \& McKee
1982), which uses the measured lag between variability in the
photoionizing continuum and emission lines to estimate BLR radii.
Currently there are 35 reverberation-mapped AGNs in the literature
(Peterson \etal\ 2004).  These, in turn, are used to calibrate the
radius-luminosity relation, \rblr $\propto$ \lf$^{0.64}$ (Greene \& Ho
2005b; see also Kaspi \etal\ 2005), from which it is possible to infer
a radius from the AGN luminosity.  
Combining the radius with an estimate of the velocity dispersion of the BLR
yields a virial mass estimate for the BH, \mbh=$fR~v^2/G$, where $f$ is
a factor that accounts for the geometry of the BLR.  We
assume $f=0.75$ for a spherical BLR (Netzer 1990).  Because our
sample is selected to have strong stellar continua, virial BH mass
estimation requires special care.  Under these circumstances the most
robust virial mass estimator is that based on the width and luminosity
of the broad \halpha\ emission line, as advocated by Greene \& Ho (2005b).
The emission-line fitting is described in Greene \& Ho
(2004, 2005b).  Briefly, the stellar continuum is modeled and removed
using a principal component analysis designed for the SDSS data by Lei Hao
(Hao \etal\ 2005).  We then construct a multi-component Gaussian model
of the \sii~$\lambda \lambda$6716, 6731 doublet, which is shifted and
scaled to fit the narrow \halpha+\nii~$\lambda\lambda$6548, 6583
complex.  The remaining (broad) \halpha\ flux is fit with as many
Gaussian components as needed to provide a statistically acceptable
fit, although we attach no physical significance to the individual
components.  As described in Greene \& Ho (2005b), we measure both
\fwha\ and the true second moment (\sigha) from the multi-component
Gaussian fits.  The uncertainties in the luminosities and line widths
are derived using simulations, although we set a minimum uncertainty
of $\sim 5 \%$ on the line width.

\section{Results}

We present the largest single collection of \sigmastar\ measurements
in active galaxies.  The single-epoch \mbh\ values for the SDSS and
Barth \etal\ (2005) samples are estimated using \fwha\ (rather
than \sigha; e.g.,~Peterson \etal\ 2004).  In our full sample, we also
include the 15 reverberation-mapped AGNs considered by either Onken
\etal\ (2004) or Nelson \etal\ (2004), using the weighted-mean
\sigmastar\ from the two works.  Weighted-mean virial products for
this sample are computed using the \hbeta\ lag and FWHM measurements
presented in Peterson \etal\ (2004; the weights are calculated taking
into account the asymmetric error bars), and BH masses
are derived assuming $f=0.75$.  Finally, we include the
well-
\hskip -8mm
\psfig{file=msigma_qn_newmass.epsi,width=9cm,angle=0} 
\vskip -1mm
\figcaption[]
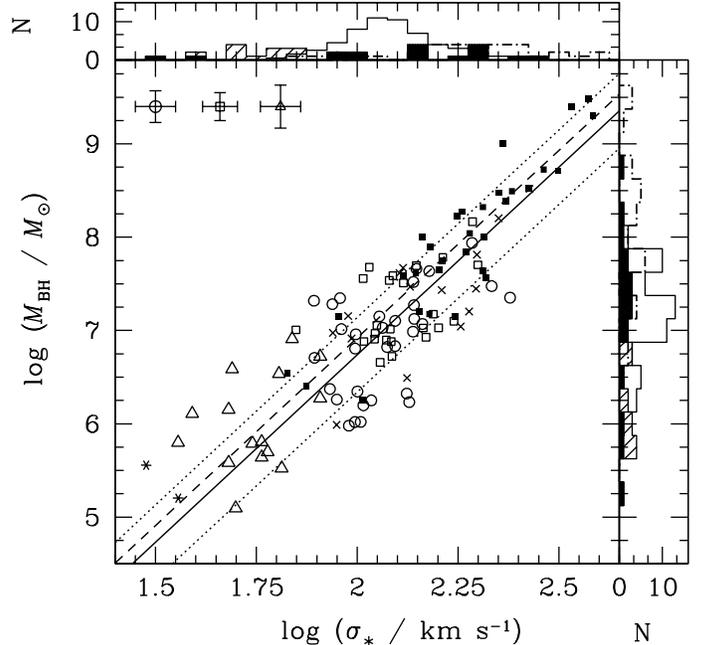
{ The \msigma\ relation for active and inactive galaxies.
Open circles are measurements from this work using \cat, open squares
are measurements using the 5250--5820 \AA\ Fe region (see Greene \& Ho
2005c), and open triangles represent the Keck data from Barth \etal\
(2005) but with \mbh\ recalculated using \fwha\ values from DR3.
Literature data include those of Onken \etal\ (2004) and Nelson \etal\
(2004; crosses; BH masses from Peterson \etal\ 2004), NGC 4395 and POX
52 (Filippenko \& Ho 2003; Peterson et al.  2005; Barth \etal\ 2004;
asterisks), and the primary sample of inactive galaxies with
dynamically determined values of \mbh\ (Tremaine et al. 2002; filled
squares).  Representative error bars are shown in the upper left for
the SDSS and Keck measurements.  The dashed line represents the fit
for the \msigma\ relation of inactive galaxies as given by Tremaine
\etal\ (2002).  The solid line is our best fit for the AGN sample with
a fixed slope of $\beta = 4.02$ and $\alpha = 7.96 \pm 0.03$; 
the dotted lines show the measured
intrinsic scatter of 0.4 dex.  We display histograms of
the \sigmastar\ ({\it top}) and \mbh\ ({\it right}) distributions for
the Keck (shaded), SDSS (open), literature (filled), and inactive
(dash-dot line) samples.
\label{msigma}}
\vskip 2mm 
\noindent
studied intermediate-mass BHs in NGC 4395 (reverberation mass
\mbh$= 3.6 \times 10^5$~\msun, Peterson \etal\ 2005; virial mass
\mbh$=1.4 \times 10^4$~\msun, Filippenko \& Ho 2003) and in POX 52
(virial mass \mbh$=1.6 \times 10^5$~\msun, Barth \etal\ 2004).  
The full sample of 88 objects 
covers a range in mass of $10^{5}-10^{8.5}$ \msun, and a
corresponding range in \sigmastar\ of $30 - 268$ \kms\ (see
Table 1).  Consistent with
previous work, Figure 1 shows that active galaxies apparently follow
the \msigma\ relation defined by inactive systems, even with a sample
size increased by a factor of nearly 6.  In detail, however, we do
find significant differences in zeropoint and slope compared to the
inactive sample, which we describe below.

Assuming a log-linear form for the \msigma\ relation, 
log(\mbh/\msun) = $\alpha+\beta$log(\sigmastar/$\sigma_0$), with
$\sigma_0$ = 200 \kms, we can formally quantify deviations of AGNs
from the \msigma\ relation of inactive systems.
We use the Levenberg-Marquardt algorithm as implemented by 
{\it mpfit} to minimize \chisq, defined, following Tremaine et al. (2002), as

\begin{equation}
\chisq \equiv \sum_{i=1}^N \frac{(y_i-\alpha-\beta
  x_i)^2}{\epsilon^2_{yi}+\beta^2\epsilon^2_{xi}},  
\end{equation}
where $x_i$ correspond to \sigmastar$_{,i}$, $y_i$ to \mbh$_{,i}$, and
$\epsilon_i$ to the formal uncertainties in each measurement.  In
order to account for asymmetric uncertainties in each parameter, at
each iteration $\epsilon_i$ is selected to reflect the sign of
$y_i-\alpha-\beta x_i$.  In the simplified case of symmetric errors,
we recover the results of the Numerical Recipes routine {\it fitexy}
(Press et al. 1992) as implemented in IDL.  We begin by fixing the
slope $\beta$ to the best-fit value of 4.02 from Tremaine et al.  and
investigate potential offsets in zeropoint.  We find, for our sample
alone, a best-fit $\alpha = 7.96 \pm 0.03$, which corresponds to an
offset of $-0.17 \pm 0.07$ from the value of $\alpha=8.13 \pm 0.06$ of
Tremaine \etal\ (consistent with Gebhardt \etal\ 2000b; Nelson \etal\
2004; Onken \etal\ 2004; see summary in Table 2).  
The offset increases to $-0.21 \pm 0.06$
when the literature values are included (using either the
reverberation-mapped or virial mass for NGC 4395).  Note that we are
using formal uncertainties in \mbh, while the true uncertainties are
probably dominated by our ignorance of the BLR geometry and the
corresponding uncertainty in how to extract the velocity dispersion of
the virialized BLR from the observed line profile.  
Following Tremaine \etal\ and Gebhardt \etal\ (2000a), we
estimate the intrinsic scatter as the value $\epsilon_0$ that, when
added to $\epsilon_y$, results in a minimum \chisqr\ of 1.  For both
our sample alone and including the literature sample, we find an
intrinsic scatter\footnote{The intrinsic scatter increases to 0.5 dex
when \sigha, rather than \fwha\ is used to estimate \mbh, partly
because of the difficulty of measuring the low-contrast line wings.
We therefore consider only masses derived using \fwha\ in the
following. We estimate that disk rotation contributes at most a small
error to \sigmastar\ since we find no correlation between galaxy
inclination angle and deviation from \msigma\ ($78 \%$ probability of
no correlation with Kendall's $\tau$), and typically only the inner
$\sim 20 \%$ of the galaxy light enters the fiber.  See further
arguments in Greene \& Ho (2005a).} of \epsi=0.4 (dotted lines in
Fig. 1).  This is to be compared with the estimated scatter of 0.25-0.3
dex for the inactive sample (Tremaine \etal\ 2002).
While there is probably intrinsic scatter in the underlying
\msigma\ relation, we conservatively adopt 0.4 dex as the systematic
uncertainty in the single-epoch \mbh\ measurements, so that we bracket
the full range in possible uncertainty.
If we repeat the fitting above increasing the uncertainties in the
single-epoch BH masses by 0.4 dex in quadrature, the results for our sample
alone are virtually unchanged, while we find $\alpha = 7.86 \pm 0.04$ for the
full sample, which corresponds to an offset of $-0.27 \pm 0.07$.  This
offset is similar to that advocated by Onken et al., but it is
statistically much more robust, being based on a final sample that is
nearly 6 times larger.  

When we fit both the zeropoint and slope, we obtain $\alpha=7.84 \pm
0.04$ and $\beta=3.65 \pm 0.13$ ($\alpha=7.89 \pm 0.05$, $\beta=3.74
\pm 0.17$) if we include (exclude) literature values, corresponding to
slopes flatter than the value of $\beta = 4.02 \pm 0.32$ for inactive
sources by $-0.37 \pm 0.35$ ($-0.28 \pm 0.36$).  Including the 0.4 dex
scatter to the single-epoch measurements yields an even flatter slope,
with $\alpha=7.79 \pm 0.04$ and $\beta = 3.49 \pm 0.18$ ($\alpha=7.68
\pm 0.10$, $\beta= 2.95 \pm 0.32$).  
It appears that much of the flattening is driven
by objects with BH masses $< 10^6$~\msun, which rely on an
extrapolation of the radius-luminosity relation and thus may be
suspect. [As pointed out by Barth \etal\ (2005), these objects also
may be biased by selection toward large \mbh\ at a given \sigmastar.]
When we remove the Barth \etal\ objects, as well as POX 52 and NGC
4395, but include all other data, we find $\beta = 4.19 \pm 0.22$ with
the formal errors, or $\beta = 3.85 \pm 0.25$ with the additional 0.4
dex uncertainty added. The fact that the slope is still shallower when
the (more realistic) uncertainties are included gives us some
confidence that the flattening is real.

\section{Implications and Conclusions}

We have established, with statistical confidence for the first time,
that the \msigma\ relation of local AGNs, while generally similar to
that of inactive galaxies, shows some significant differences.  We
find evidence for a lower zeropoint, shallower slope, and (probably)
larger scatter.  There are many competing effects,
relating both to BLR physics and galaxy evolution, 

\hskip 10mm
\psfig{file=table2_v2.epsi,width=6cm,angle=0} 
\vskip +3mm 
\noindent
that may contribute
to the observed differences.  If we posit that AGNs obey the relation
of inactive galaxies exactly, then the scatter and zeropoint offset
may be attributed to variations in the geometry of the BLR. This is
the assumption made by Onken \etal\ (2004), who derive a statistical
offset in the factor $f$ that scales all BH masses upwards relative to
the case of a spherical BLR. If, however, as suggested by disk-wind
models (e.g.,~Proga \& Kallman 2004), the geometry of the BLR depends
on physical properties of the AGN such as the BH mass and the
Eddington ratio, then virial mass estimates will not scatter randomly
about one fixed value of $f$.  Rather, $f$ will change systematically
with the state of the system, and its average value for any given
sample will depend on the range of parameter space spanned by the
objects used to derive it.  It is important to recognize that all work
so far---including ours---has considered a rather limited range in
\mbh\ and \lledd, and so may lead to a biased value of $f$. Apart from
the Greene \& Ho (2004) objects studied by Barth \etal\ (2005), the
practical challenge of detecting stellar absorption features to
measure \sigmastar\ inevitably biases the final sample toward
relatively low Eddington ratios and BH masses.  Excluding the Barth et
al. objects, the sample summarized in Figure 1 has a median \lledd\ =
$6\times10^{-2}$ and $10^6$ \lax\ \mbh\ \lax\ $10^8$~\msun.  While we
expect that secondary parameters (e.g., \lledd) may ultimately help to
account for the overall scatter in the AGN \msigma\ relation, we
refrain from discussing this issue here because we believe that a
proper analysis would require a larger and more complete sample than
is currently available.

Apart from systematic uncertainties in virial masses, evolution of the
\msigma\ relation with cosmic time (e.g.,~Shields \etal\ 2003; Treu
\etal\ 2004; Walter \etal\ 2004; Peng \etal\ 2005) also imprints
scatter and skewness into the local relation as a function of \mbh\
(Robertson \etal\ 2005).  However, it is unclear at this stage how
much scatter can be attributed to the fact that BHs in AGNs, in so far
as they are radiating and thus accreting, are still gaining mass.  
Any differential growth between BHs and bulges
will introduce additional scatter to the AGN \msigma\ relation
if AGN accretion and star formation are not precisely synchronized (Ho
2005; Kim et al. 2005).  If we
ascribe all the observed scatter to differential growth (taking the
observed scatter of $\sim 0.27$ dex in the relation for inactive
objects from Tremaine \etal), there can be no more than $\sim$0.3 dex
scatter (factor of 2) introduced by relative BH-bulge growth.  
This level of growth has only modest fuel requirements that are easy
to sustain in nearby galaxies; 
a $10^6$ \msun\ BH requires only
$0.022$~\msun\ yr$^{-1}$ to double its mass in a Saltpeter time ($4.5
\times 10^{7}$ yr).

Finally, taking our entire data set at face value, we do find evidence
for a shallower slope than the inactive \msigma\ relation.  Whether
this is the result of flattening at low mass or a different slope in
AGNs is difficult to determine at the present time, in the absence of
more AGNs with \mbh\ $> 10^8$~\msun.  In this regard, Wyithe (2005)
argues that the slope for the inactive sample steepens considerably
when the four smallest \sigmastar\ values are removed, independently
suggesting flattening at low mass.  Flattening at low mass may be a
result of the changing efficiency in AGN feedback in low-mass halos.
For instance, the formalism of Vittorini \etal\ (2005) finds that a
combination of decreased optical depth (proportional to galaxy radius)
and cooling times shorter than a dynamical time (small, dense halos)
allows these BHs to grow to larger relative masses without feedback
limitations.  Alternatively, different growth modes between spheroidal
and disk-dominated systems may result in a different final position on
the \msigma\ plane.  On the other hand, we cannot preclude the
possibility that a change in slope in the BLR radius-luminosity
relation at low luminosity is responsible for the change in slope in
the \msigma\ relation.  However, we note that a radius-luminosity
slope closer to the theoretically preferred value of 0.5 (e.g.,~Kaspi
\etal\ 2000) would only increase the observed discrepancy in slope.
Reverberation mapping of low-mass BHs would be required to address
this issue.

In summary, although we have increased the population of AGNs with \sigmastar\ 
measurements by a factor of nearly 6, we do not yet have a large enough 
sample, or adequate coverage of parameter space, to uniquely identify the 
cause or causes responsible for the observed differences between 
the \msigma\ relation of local active and inactive galaxies.  Future effort 
should focus on (1) further enlarging the sample of AGNs with robust 
\sigmastar\ measurements, (2) pushing the samples to the extremes of the 
mass distribution, particular above $10^8$~\msun, (3) extending the 
luminosity coverage to include objects with a wider range in Eddington 
ratios, and (4) better characterizing the BLR radius-luminosity relation 
over a broader range of AGN properties than is currently available.

\acknowledgements 
We thank L. Hao for providing her PCA code, D.~Proga
for useful discussion, and the SDSS collaboration for providing the
extraordinary database and processing tools that made this work
possible.  We thank an anonymous referee for a prompt and positive
review.


\end{document}